\documentclass[prb,twocolumn,showpacs,preprintnumbers,amsmath,amssymb]{revtex4}
\usepackage{graphicx}
\usepackage{dcolumn}
\usepackage{bm}

\begin{document}

\title{Itinerant antiferromagnetism in BaCr$_2$As$_2$}

\author{D.J. Singh}
\author{A.S. Sefat}
\author{M.A. McGuire}
\author{B.C. Sales}
\author{D. Mandrus}
\affiliation{Materials Science and Technology Division,
Oak Ridge National Laboratory, Oak Ridge, Tennessee 37831-6114}

\author{L.H. VanBebber}
\author{V. Keppens}
\affiliation{Department of Materials Science and Engineering,
The University of Tennessee, Knoxville, TN 37996-2200}

\date{\today}

\begin{abstract}
We report single crystal synthesis,
specific heat and resistivity measurements and electronic
structure calculations for BaCr$_2$As$_2$. This material is a
metal with itinerant antiferromagnetism, similar to the parent phases of
Fe-based high temperature superconductors, but differs in magnetic
order.
Comparison of bare band structure density of states and the low temperature
specific heat implies a mass renormalization of $\sim$ 2.
BaCr$_2$As$_2$ shows stronger transition
metal - pnictogen covalency than the Fe compounds, and in this respect
is more similar to BaMn$_2$As$_2$. This provides an explanation for
the observation that Ni and Co doping is effective in the Fe-based
superconductors, but Cr or Mn doping is not.
\end{abstract}


\maketitle

\section{introduction}

The discovery of high temperature Fe-based superconductivity
\cite{kamihara-a}
has resulted in substantial activity leading to the finding of
a wide range of Fe-based superconductors, including a number
of materials in the ThCr$_2$Si$_2$ structure, prototype BaFe$_2$As$_2$.
\cite{rotter-sc}
Similar to the oxy-arsenides, BaFe$_2$As$_2$ is metallic and shows
spin density wave (SDW) antiferromagnetism when cooled.
\cite{rotter-sdw}
BaFe$_2$As$_2$ becomes superconducting when the SDW is destroyed,
either by doping with holes \cite{rotter-sc}
or electrons, \cite{sefat-co}
or by using pressure. \cite{alireza}
In fact, it is remarkable that superconductivity can be induced
by doping on the Fe-site, using Co and Ni as electron dopants.
\cite{sefat-co,matsuishi}
This is in contrast to the behavior observed in cuprate superconductors,
where alloying on the Cu site strongly suppresses superconductivity.

The electronic structure of BaFe$_2$As$_2$ is similar to the
other iron based superconducting materials in that the
Fermi energy ($E_F$) lies in at the
bottom of a pseudogap in the electronic density of states. The
corresponding Fermi surface then consists of small disconnected
hole and electron pockets around the zone center and zone corner.
\cite{singh-bfa,nekrasov,singh-du,mazin}
Importantly the bands within 2 eV of the Fermi energy arise from
Fe $d$ states with only modest As $p$
hybridization.
BaCo$_2$As$_2$ and BaNi$_2$As$_2$, show very different properties,
namely those of a material very close to ferromagnetism,
\cite{sefat-baco}
and a low temperature electron-phonon superconductor, respectively.
\cite{ronning-banias,kurita-banias,subedi-banias}
Importantly, however, the electronic structures of Ba$T_2$As$_2$, $T$=Fe,Co,Ni,
are closely related, with Fe $d$ bands near $E_F$, modest
hybridization with As, and a similar shaped density of states with
a pseudogap at an electron count of six $d$ electrons. The main
differences in physical properties are due to the different electron
counts of Fe$^{2+}$, Co$^{2+}$ and Ni$^{2+}$. This common behavior
also is thought to be important for the superconductivity in the
Co-doped superconductor BaFe$_{2-x}$Co$_x$As$_2$, whose electronic
structure behaves as a coherent alloy. \cite{sefat-co}
On the other hand,
to our knowledge, superconductivity has not been reported in
BaFe$_2$As$_2$ doped with Mn or Cr so far.
BaMn$_2$As$_2$ shows an electronic structure very different from
the $T$=Fe,Co,Ni compounds.
In particular, it has strong spin dependent hybridization between
Mn $d$ states and As $p$ states, and is a small band gap
semiconductor with a high exchange couplings
and ordering temperature that result from the strong hybridization.
\cite{an,ysingh}
Here we report investigation of the $T$=Cr material, BaCr$_2$As$_2$.
This material is known to form in the ThCr$_2$Si$_2$ structure,
\cite{pfisterer}
but little has been reported about its physical properties.
We find that similar to BaMn$_2$As$_2$ it has strong spin dependent
Cr $d$ - As $p$ hybridization, but unlike that material it is a
renormalized antiferromagnetic metal, with a $G$-type (checkerboard)
order and very strong magnetic interactions.

\section{methods}

BaCr$_2$As$_2$ single crystals were prepared starting from
high purity elements ($>$ 99.9\%, source Alfa Aesar).
The crystals were grown out of CrAs binary.
Cr powder and As pieces were reacted slowly by heating to 300 $^\circ$C
(50 $^\circ$C/hr, dwell 10 hrs), to 600 $^\circ$C (30 $^\circ$C/hr
dwell 30 hrs), then to 900 $^\circ$C (30 $^\circ$C/hr, dwell 24 hrs).
A ratio of Ba:CrAs = 1:4 was heated for 13 hours at 1230 $^\circ$C under
partial argon atmosphere. The ampule was cooled at the rate of 2 $^\circ$C/hr,
followed by decanting of flux at 1120 $^\circ$C.
Electron probe microanalysis of a cleaved surface of the single crystal
was performed on a JEOL JSM-840 scanning electron microscope using
an accelerating voltage of 15 kV and a current of 20 nA with an EDAX
brand energy-dispersive X-ray spectroscopy (EDS) device. EDS analyses on
the crystal indicated a Ba:Cr:As ratio of 1:2:2, within the error bars.
The phase purity of the crystals was determined using a Scintag XDS 2000
2$\Theta$-2$\Theta$ diffractometer (Cu K$_\alpha$ radiation).
BaCr$_2$As$_2$ crystallizes with the ThCr$_2$Si$_2$ structure at room
temperature (tetragonal spacegroup $I4/mmm$, No. 139, $Z$=2).
Lattice constants were determined from LeBail refinements using the
program FullProf. \cite{fullprof} The resulting room
temperature lattice parameters were $a$=3.9678(4) \AA, $c$= 13.632(3) \AA.

Temperature dependent electrical resistivity measurements were performed
on a Quantum Design Physical Property Measurement System (PPMS). The
electrical contacts were placed on the samples in the standard 4-probe
geometry, using Pt wires and silver paste. The resistivity
was measured in the $ab$-plane, i.e. $\rho_{ab}(T)$.
Specific heat data, $C_p(T)$, were also obtained using the PPMS.
The relaxation method was used from 2 K to 200 K.

The first principles
calculations were done within the local density approximation
(LDA) using the general potential linearized augmented planewave (LAPW)
method, \cite{singh-book}
similar to prior calculations for BaFe$_2$As$_2$. \cite{singh-bfa}
We used the reported experimental lattice parameters from literature,
\cite{pfisterer}
$a$=3.963\AA, $c$=13.600\AA,
which are very close to the room temperature values determined here.
The internal parameter, $z_{\rm As}$ was determined by
energy minimization.
The resulting value for the lowest energy G-type antiferromagnetic
ordering is $z_{\rm As}$=0.3572.
This value shows sensitivity to magnetic order as in the
Fe-based compounds. \cite{mazin-mag}
A relaxation for ferromagnetic order
yielded $z_{\rm As}$=0.3526.
We used well converged basis sets, including local orbitals to
treat the semi-core states and relax the Cr $d$ state linearization.
\cite{singh-lo}
Relativistic effects were included at the scalar relativistic level.
The LAPW sphere radii were 2.2 $a_0$ for Ba and 2.1 $a_0$ for Cr and As.

\section{density functional calculations}

We begin with a discussion of the magnetic order. We calculated the
energy as a function of magnetic ordering for different likely configurations.
These were a non-spin-polarized calculation (i.e. no magnetism), and
three magnetic orderings of the Fe planes, with both ferromagnetic and
antiferromagnetic layer stackings for each.
These orders were ferromagnetic, checkerboard nearest neighbor
antiferromagnetism and a magnetic structure consisting of ferromagnetic
chains of nearest neighbor Fe, alternating antiferromagnetically, as in
the SDW state of the undoped Fe superconducting materials.
The results are summarized in Table \ref{tab-mag}.
As may be seen, checkerboard order with antiferromagnetic
stacking, i.e. G-type antiferromagnetism, yields the lowest energy.
This state has a Cr moment of $\sim$ 2 $\mu_B$.

\begin{table}[tbp]
\caption{LDA magnetic energy, $E_{mag}$ of BaCr$_2$As$_2$. The energies are
given for $z_{\rm As}$=0.3572, on a per formula unit basis (two Cr atoms)
with respect to the non-spin-polarized energy. The moments, $m_{\rm Cr}$
are the spin moments defined by the integral over the Cr LAPW sphere.
The notation for the magnetic order is as follows:
P denotes non-spin-polarized, otherwise the first letter denotes
the in-plane magnetic order (F for ferromagnetic, C for checkerboard
nearest neighbor antiferromagnetism, and S for chains of like-spin Fe,
as in the SDW of the Fe-based compounds), and the second denotes either
antiferromagnetic stacking (A) or ferromagnetic stacking (F) along the
$c$-axis.
}
\vspace{0.25cm}
\label{tab-mag}
\begin{tabular}{|l|c|c|}
\colrule
 ~Magnetic Order~   &  ~~~$E_{mag}$ (eV)~~~   & ~~~~$m_{\rm Cr}$ ($\mu_B$)~~~~ \\
\colrule
P   &    0.000  &  0.00 \\
F-F &   -0.157  &  1.56 \\
F-A &   -0.146  &  1.51 \\
S-F &   -0.059  &  1.73 \\
S-A &   -0.058  &  1.80 \\
C-F &   -0.382  &  2.05 \\
C-A &   -0.394  &  2.01 \\
\colrule
\end{tabular}
\vspace{0.25cm}
\end{table}

Turning to the details, one may note that although the magnetic
energy is large, $\sim$ -0.2 eV / Cr for the ground state,
and all the magnetic ordering patterns tested are lower energy than
the non-spin-polarized case,
the differences between different orderings are of almost the
same scale. For example, the difference between the chain-like
order (S-A) that is the ground state for the Fe-based superconductors
and the G-type ground state of BaCr$_2$As$_2$ is 0.17 eV/Cr,
i.e. $\sim$ 85\% of the magnetic energy in the ground state.
This implies that the magnetism has significant itinerant character,
in the sense that band structure (hopping) is important
in the moment formation, as opposed to just atomic physics, with
hopping important only in the inter-site exchange.
Also, one may note that the $c$-axis coupling is small compared to the
in-plane coupling, and is antiferromagnetic for the ground state, but
that the sign depends on the details of the in-plane order, implying that
more than nearest neighbor interactions are important. This is also
apparent from the in-plane energetics. A simple fit of the energies
for antiferromagnetic stacking to a
Heisenberg model with nearest neighbor ($J_1$)
and next nearest neighbor ($J_2$) interactions, would yield antiferromagnetic
$J_1$ and a large opposite (ferromagnetic)
$J_2$, $J_2$/$J_1$=-0.85. The large magnitude of
$J_2$ implies that this model is probably not reliable and that the
interactions are probably long range, as might be expected in a metal
with itinerant magnetism.

\begin{figure}[tbp]
\vspace{0.25cm}
\includegraphics[height=0.96\columnwidth,angle=270]{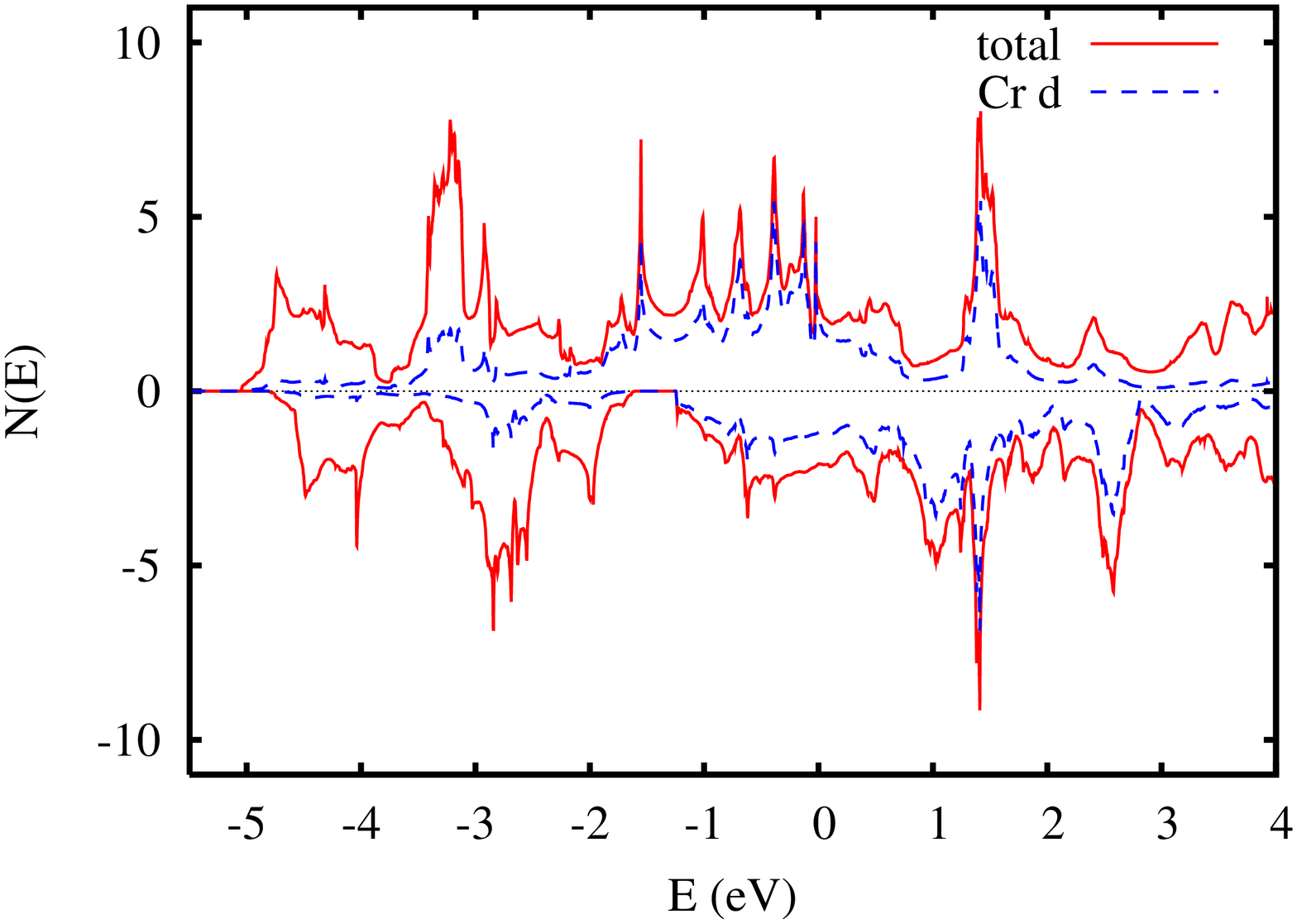}
\includegraphics[height=0.96\columnwidth,angle=270]{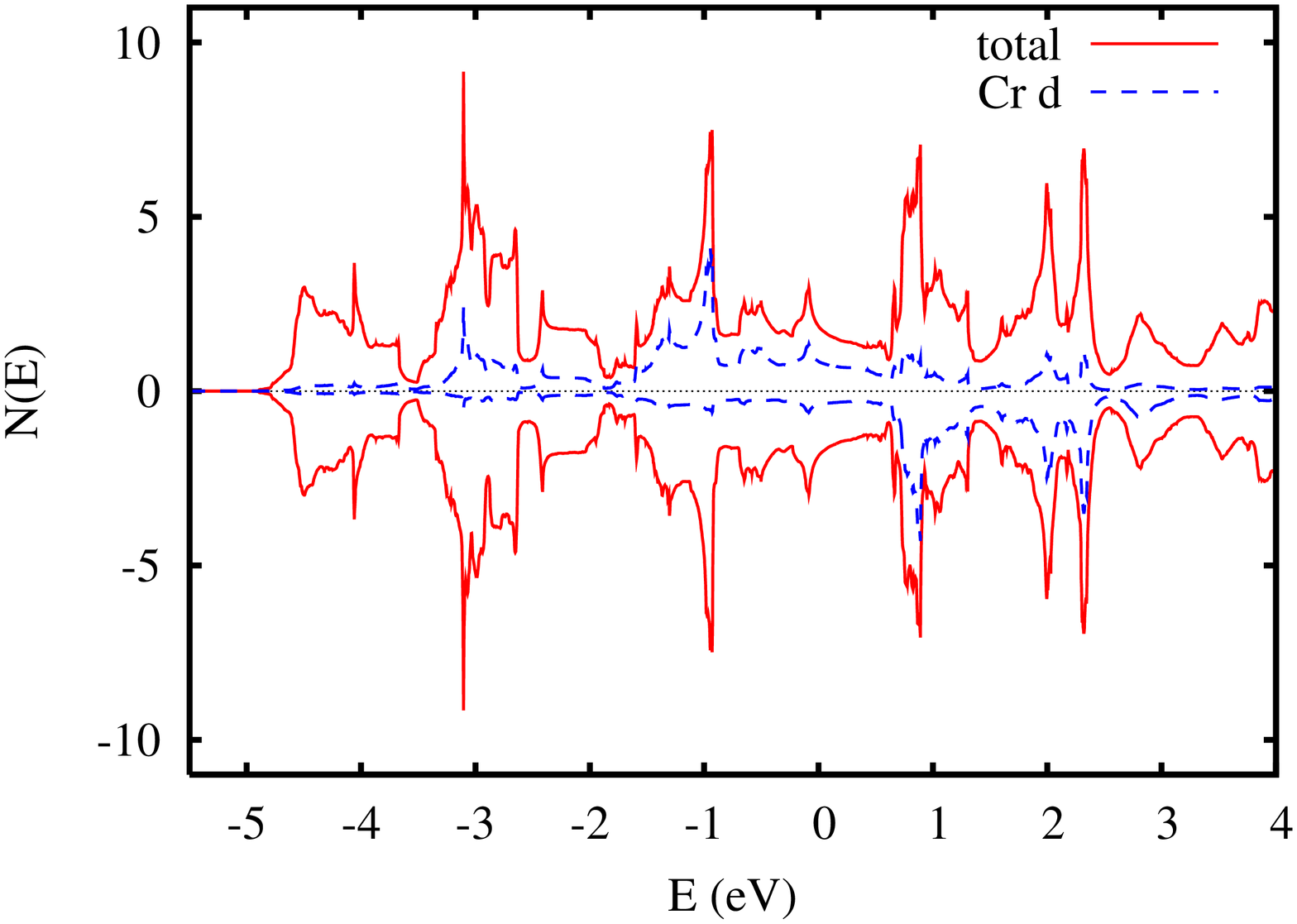}
\caption{Calculated electronic
density of states BaCr$_2$As$_2$ with ferromagnetic (top) and
nearest neighbor antiferromagnetic (bottom) ordering. Majority spin
is shown above the axis and minority spin below. The
projection is onto the LAPW sphere, radius 2.1 $a_0$.
In the top panel the Cr projections are for both atoms. In the bottom
panel, spin up and spin down are identical, and the projections
shown are majority and minority spin for one Cr atom. The total DOS is
per formula unit.}
\label{dos}
\end{figure}

\begin{figure}[tbp]
\vspace{0.25cm}
\includegraphics[height=0.96\columnwidth,angle=270]{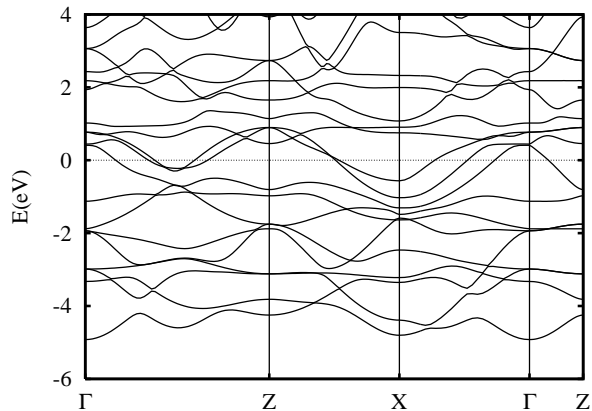}
\caption{Calculated band structure of BaCr$_2$As$_2$ with nearest
neighbor antiferromagnetic ordering.}
\label{bands}
\end{figure}

\begin{figure}[tbp]
\vspace{0.25cm}
\includegraphics[width=0.99\columnwidth,angle=0]{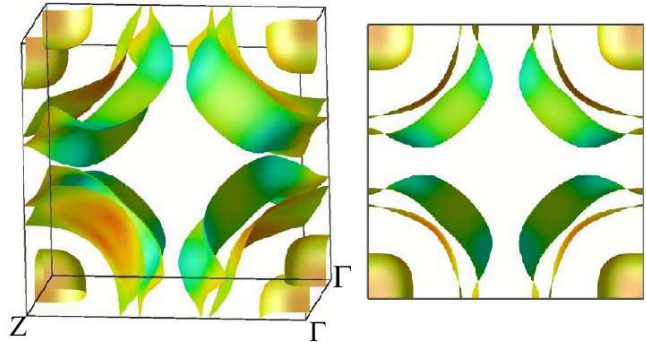}
\caption{(color online)
Calculated extended zone
Fermi surface of antiferromagnetic BaCr$_2$As$_2$,
showing a view off the $c$-axis (left) and along the $c$-axis (right).
The shading is by velocity.}
\label{fermi}
\end{figure}

The main results for the LDA electronic structure are given in Figs.
\ref{dos}, \ref{bands} and \ref{fermi}, which show
the electronic density of states (DOS), the band structure and
the Fermi surface, respectively for the G-type ground state order.
Fig. \ref{dos} shows in addition the DOS for a ferromagnetic order.
The compound is metallic for either order.

An examination of the DOS shows that the states near the $E_F$
are hybridized Cr $d$ - As $p$ combinations and that the hybridization
is spin-dependent. This is similar to BaMn$_2$As$_2$, but rather
different from the $T$=Fe,Co,Ni series where, as mentioned, the
states near $E_F$ are dominated by $d$ character.
In analyzing the DOS, it is important to keep in mind that the projections
in the LAPW method are onto the LAPW spheres. Since a Cr $3d$ orbital
is almost entirely contained within a 2.1 $a_0$ sphere, the Cr $d$ projection
is a reasonable approximation to the Cr $d$ contribution to the DOS. However,
As $p$ states are extended and would have substantial weight outside a
sphere of this radius. Thus when considering the electronic structure,
the difference between the total DOS and the Cr $d$ projection is a better
measure of the As $p$ contribution than the As $p$ projection, which would
underestimate the As contribution. Viewed in this way, As $p$ orbitals
contribute approximately 1/3 of the DOS at $E_F$ and furthermore
of the remaining 2/3, which is Cr $d$ in character, the majority spin
contributes more than twice as much as the minority spin.
As may be seen comparing the top and bottom panels of Fig. \ref{dos},
the details of this spin dependent hybridization are sensitive to the
magnetic order. This explains the large energy differences between
different magnetic orders.

As mentioned, BaCr$_2$As$_2$ is metallic. We find a large multisheet
Fermi surface. This consists of small rounded electron cubes around
$\Gamma$, and two electron cylinders, also centered at $\Gamma$ and
running along the $k_z$ direction. The outermost cylinder in particular
has significant corrugation.

The density of states at the Fermi energy is moderately high,
$N(E_F)$=3.7 eV$^{-1}$ on a per formula unit, both spins basis.
This corresponds to an non-renormalized bare band specific heat coefficient
$\gamma_0$=9.3 mJ/(K$^2$ mol).
The calculated anisotropy is modest.
The $ab$-plane and $c$-axis Fermi velocities are
$<v_x^2>^{1/2}$=2.08x10$^5$ m/s and
$<v_z^2>^{1/2}$=1.08x10$^5$ m/s, respectively.
If the scattering rate is isotropic this would correspond
to $\rho_{c}$/$\rho_{ab}$=4. If the
scattering rate is the same for all Fermi surfaces,
$c$-axis conduction will be
roughly equally from the small $\Gamma$ centered pocket and the
outer electron cylinder, while the in-plane conduction will come
mostly from two cylinders.

\section{transport and specific heat}

\begin{figure}[tbp]
\vspace{0.25cm}
\includegraphics[width=\columnwidth,angle=0]{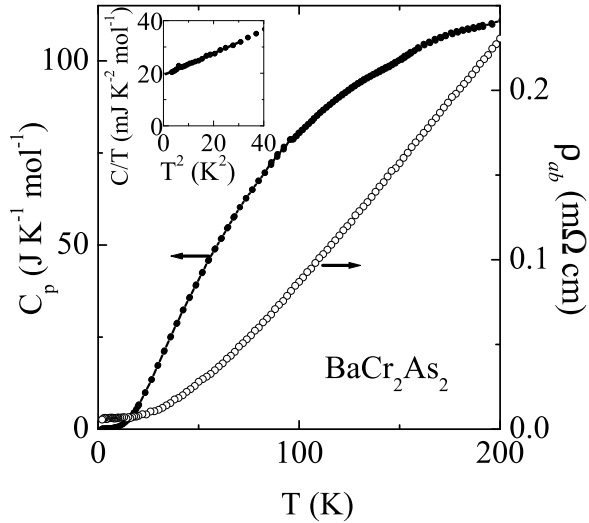}
\caption{
Single crystal specific heat and resistivity
of BaCr$_2$As$_2$ as a function of temperature. The inset shows
the linear dependence of $C/T$ vs. $T^2$ with finite value at $T=0$.}
\label{cp}
\end{figure}

Fig. \ref{cp} shows the temperature dependent resistivity and specific
heat for a single
crystal of BaCr$_2$As$_2$. As mentioned, the
resistivity was measured in the $ab$ plane.
As may be seen, the sample is clearly metallic with an
increasing $\rho(T)$ and a finite $C_P(T)/T$ at low $T$.
This is in accord with the results of our LDA calculations.
There are no features suggesting a phase transition in either
$\rho(T)$ or $C_P(T)$ from 2 K to 200 K.
This is perhaps not surprising considering the large energy
differences between different magnetic configurations found in the
LDA calculations, as these differences would suggest an ordering
temperature well above the maximum temperature of our measurements.
We find specific heat
$\gamma$=19.3(1) mJ/(K$^2$ mol), i.e. $\sim$ 9.6 mJ/(K$^2$ mol Cr).
Comparing with the bare band structure value, we find an enhancement of
$\gamma / \gamma_0 \sim$ 2.
This is a substantial renormalization, comparable to the Fe-based
superconductors, although the physics in BaCr$_2$As$_2$ may be different.
The renormalization may come in part from the electron-phonon interaction
and in part from spin-fluctations or other correlation effects.

\section{summary and conclusions}

We find that
BaCr$_2$As$_2$ is an itinerant antiferromagnetic metal with a complex
multisheet Fermi surface and substantial ($\sim$ 2) specific heat
renormalization.
While the electronic structure and magnetic order are different from
those in the Fe-based superconducting materials, these results to
suggest that BaCr$_2$As$_2$ is an interesting material for further
investigation. We also note that the stronger Cr-As covalency relative
to the Fe-based superconductors means that Cr dopant atoms
in those materials will produce more scattering than Co or Ni dopants,
perhaps explaining why superconductivity has not yet been observed in
Cr-doped BaFe$_2$As$_2$.
Finally, we note that BaMn$_2$As$_2$ and BaMn$_2$Sb$_2$ have been
discussed as potential thermoelectric materials.
\cite{an,wang}
The present results showing similar bonding with As and spin dependent
metal - As bonding in the Cr compound as was found previously in the
Mn compounds suggest that Cr could be used for doping studies of the
thermoelectric properties of the Mn phases.

\acknowledgments

This work was supported by the Department of Energy,
Division of Materials Sciences and Engineering and
by the ORNL LDRD program.

\bibliography{BaCr2As2}
\end{document}